# Correlation-corrected band topology and topological surface states in iron-based superconductors

Xiaobo Ma,[#] Guangwei Wang,[#] Rui Liu, Tianye Yu, Yiran Peng, Pengyu Zheng and Zhiping Yin*

*Department of Physics and Center for Advanced Quantum Studies, Beijing Normal BUniversity, Beijing 100875, China*

[#]These authors contributed equally to this work.
*yinzhiping@bnu.edu.cn

Iron-based superconductors offer an ideal platform for studying topological superconductivity and Majorana fermions. In this paper, we carry out a comprehensive study of the band topology and topological surface states of a number of iron-based superconductors using a combination of density functional theory (DFT) and dynamical mean field theory. We find that the strong electronic correlation of Fe $3d$ electrons plays a crucial role in determining the band topology and topological surface states of iron-based superconductors. Electronic correlation not only strongly renormalizes the bandwidth of Fe $3d$ electrons, but also shifts the band positions of both Fe $3d$ and As/Se $p$ electrons. As a result, electronic correlation moves the DFT-calculated topological surface states of many iron-based superconductors much closer to the Fermi level, which is crucial for realizing topological superconducting surface states and observing Majorana zero modes as well as achieving practical applications, such as quantum computation. More importantly, electronic correlation can change the band topology and make some iron-based superconductors topologically nontrivial with topological surface states whereas they have trivial band topology and no topological surface states in DFT calculations. Our paper demonstrates that it is important to take into account electronic correlation effects in order to accurately determine the band topology and topological surface states of iron-based superconductors and other strongly correlated materials.

## I. INTRODUCTION

In recent years, topological insulators (TIs) [1-5] and unconventional high-temperature iron-based superconductors [6,7] have been two important research areas in condensed-matter physics. The combination of topological properties and high-temperature superconductivity has also attracted great interest. It is well known that the combination of topological properties and superconductivity offers a natural platform for realizing topological superconductivity and Majorana zero modes [8-17] utilized in topological quantum computation [18,19]. A straightforward method to integrate the topological properties and superconductivity is through a proximity effect by making TI-superconductor heterostructures [8,20,21]. However, materials [11-17,22-35] with both topological properties and superconductivity avoid potential material incompatibility, hence, are more desirable.

Nontrivial topological phases have been theoretically predicted in the single-layer FeSe [22], the monolayer and the thin-film $FeTe_{1-x}Se_x$ [26]. By angle-resolved photoelectron spectroscopy (ARPES) measurements, the topologically nontrivial band inversion has been investigated in $FeTe_{1-x}Se_x/SrTiO_3$ monolayers [27]. Using density functional theory (DFT) calculations and ARPES measurements, the topological surface states have been confirmed in Fe(Te,Se) [23,25,29] and Li(Fe,Co)As [30]. Recently, the Majorana zero modes on the surface of iron-based superconductors $FeTe_{0.55}Se_{0.45}$ [12,13,15] and LiFeAs [17] have been discovered by scanning tunneling microscopy (STM). In addition, the topological surface states and Majorana zero modes in $(Li_{0.84}Fe_{0.16})OHFeSe$ [11,36] and $CaKFe_4As_4$ [16] have also been confirmed by ARPES and STM, respectively. These results indicate that a broad class of iron-based superconductors are promising platforms for realizing topological superconductivity and Majorana zero modes.

It is well known that there is strong electronic correlation among Fe $3d$ electrons in iron-based superconductors. A combination of density functional theory and dynamical mean field theory (DFT+DMFT) has been proved to be a more adequate method to study the electronic structures of iron-based superconductors [37-39]. For

($Li_{0.84}Fe_{0.16}$)OHFeSe [11] and $CaKFe_4As_4$ [16], DFT+DMFT calculations are consistent with the ARPES results. However, DFT+DMFT calculations of the band topology and topological surface states in other iron-based superconductors have not been systematically studied yet.

In this paper, we use DFT+DMFT to comprehensively study the band topology and topological surface states of iron-based superconductors. We find that the strong electronic correlation of Fe 3$d$ electrons can change the band topology of some iron-based superconductors, such as CaFeAsF and electron doped LaFeAsO, transforming the DFT-predicted trivial band topology to nontrivial band topology, thus, inducing topological surface states. Moreover, for almost all the iron-based superconductors we studied, the topological surface states are moved much closer to the Fermi energy by electronic correlation. Therefore, it is crucial to include electronic correlation in order to accurately predict the band topology and topological surface states of iron-based superconductors.

## II. COMPUTATIONAL METHOD

First-principles calculations are carried out based on DFT [40,41] within the full-potential linearized-augmented plane-wave method implemented in the WIEN2K simulation package [42]. The Perdew-Burke-Ernzerhof (PBE) parametrization of the generalized gradient approximation is used for the exchange-correlation potential [43]. The plane-wave cutoff is $R_{MT}K_{\max} = 8$. We also carry out DFT calculations using the projector augmented wave pseudopotential method implemented in the Vienna *ab initio* simulation package [44,45]. The plane-wave cutoff energy is set to 500 eV. The calculated results from these two packages are in good agreement. The irreducible representations of the electronic states are obtained by using the IRVSP code [46].

To take into account the strong electronic correlation of Fe 3$d$ electrons, we employ DFT+DMFT [47-49] to calculate the correlated band structures of iron-based superconductors. The DFT part is based on WIEN2K with the PBE exchange-correlation potential [43]. A Hubbard $U = 5.0$ eV and Hund's coupling $J = 0.8$ eV are used in all the DFT+DMFT calculations, in consistent with previous studies [37-39].

The DMFT quantum impurity problem is solved using the continuous-time quantum Monte Carlo method [50,51] at a temperature of $T$ = 116 K. The spin-orbit coupling (SOC) is included in both the DFT and DFT+DMFT calculations.

The DFT tight-binding Hamiltonian is obtained by the maximally localized Wannier functions [52-54] implemented in the WANNIER90 code [55]. Based on the DFT tight-binding Hamiltonian, we compress the hopping parameters and modify the on-site energy of both the Fe $d$ and As/Se $p$ bands to match the DFT+DMFT band structure around the Fermi level through a systematic optimization procedure. Then we obtain the DFT+DMFT tight-binding Hamiltonian. The surface states are obtained through an iterative method [56,57] implemented in the WANNIERTOOLS package [58] employing both the DFT and DFT+DMFT tight-binding Hamiltonians.

All DFT and DFT+DMFT calculations are performed in the nonmagnetic and paramagnetic state, respectively. The experimental crystal structures of FeSe [59], LiFeAs [60], NaFeAs [61], LaFeAsO [6], CaFeAsF [62], BaFe$_2$As$_2$ [63], KFe$_2$As$_2$ [64], and SrFe$_2$As$_2$ [65], including the lattice parameters and internal atomic positions, are used in the calculations. For LaFeAsOH$_{0.2}$, we add 0.2 electrons per Fe to the LaFeAsO unit cell using virtual crystal approximation, and we use the experimental lattice constants of LaFeAsO [6] whereas relax the internal coordinates of As and La ions using DFT+DMFT [66]. The experimental atomic coordinates of the parent compound LaFeAsO and the optimized atomic coordinates of LaFeAsOH$_{0.2}$ are listed in Table I.

TABLE I. Experimental atomic coordinates of the parent compound LaFeAsO and the optimized atomic coordinates of LaFeAsOH$_{0.2}$.

|      |      | LaFeAsO |      |         | LaFeAsOH$_{0.2}$ |      |         |
|------|------|---------|------|---------|------------------|------|---------|
| Atom | Site | $x$     | $y$  | $z$     | $x$              | $y$  | $z$     |
| La   | 2$c$ | 0.25    | 0.25 | 0.14154 | 0.25             | 0.25 | 0.14566 |
| Fe   | 2$b$ | 0.75    | 0.25 | 0.5     | 0.75             | 0.25 | 0.5     |
| As   | 2$c$ | 0.25    | 0.25 | 0.65120 | 0.25             | 0.25 | 0.65235 |
| O    | 2$a$ | 0.75    | 0.25 | 0       | 0.75             | 0.25 | 0       |

### III. RESULTS AND DISCUSSION

Many iron-based superconductors crystallize in a primitive tetragonal crystal structure with the space group *P*4/*nmm*, such as FeSe, LiFeAs, NaFeAs, LaFeAsO, and CaFeAsF as shown in Fig. 1(a), or a body-centered tetragonal crystal structure with the space group *I*4/*mmm*, for example, BaFe$_2$As$_2$, KFe$_2$As$_2$, and SrFe$_2$As$_2$, as shown in Fig. 1(b). The first Brillouin zones (BZs) of iron-based superconductors with the space groups *P*4/*nmm* and *I*4/*mmm* are shown in Figs. 1(c) and 1(d), respectively.

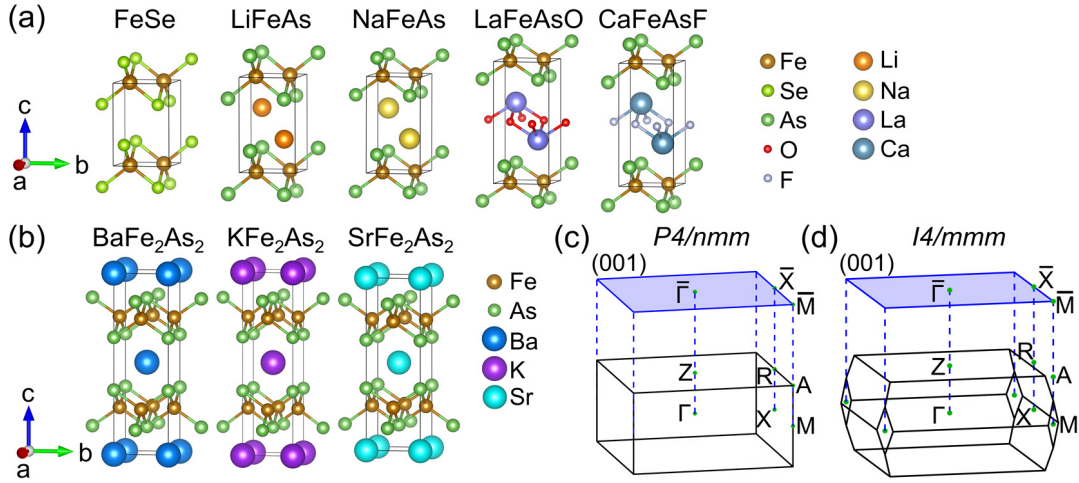

FIG. 1. Crystal structures of iron-based superconductors with the space groups (a) *P*4/*nmm* and (b) *I*4/*mmm*. Bulk Brillouin zones (BZs) and two-dimensional BZs of the projected (001) surfaces of iron-based superconductors with the space groups (c) *P*4/*nmm* and (d) *I*4/*mmm*. The reciprocal points in the bulk BZ with the space group (c) *P*4/*nmm* are Γ (0,0,0), *X* (0.5,0,0), *M* (0.5,0.5,0), *Z* (0,0,0.5), *R* (0.5,0,0.5) and *A* (0.5,0.5,0.5). The reciprocal points in the bulk BZ with the space group (d) *I*4/*mmm* are Γ (0,0,0), *X* (0.25,-0.25,0.25), *M* (0,0,0.5), *Z* (0.5,0.5,-0.5), *R* (0.75,0.25,-0.25) and *A* (0.5,0.5,0). The reciprocal points in the BZ are given as linear combinations of the reciprocal lattice vectors of the primitive cell.

**A. Electronic correlation moves the topological surface states much closer to the Fermi level**

Our DFT+DMFT calculations reveal that electronic correlation moves the topological surface states much closer to the Fermi level than DFT calculations. This effect generally applies to all the iron-based superconductors studied in this paper.

Taking LiFeAs as an example, we carry out a detailed study of the band topology and topological surface states using both DFT and DFT+DMFT. When SOC is ignored, the DFT band structure of LiFeAs along the high-symmetry lines in the first BZ is shown in Fig. 2(a). When SOC is included, the $d_{xz}/d_{yz}$ band along the Γ-Z direction splits into two $d_{xz}/d_{yz}$ bands. The inclusion of SOC avoids the crossing of the $p_z$ ($\Lambda_6$) band and the lower $d_{xz}/d_{yz}$ ($\Lambda_6$) band and opens a SOC gap between them, hence, the band structure has a continuous direct gap in the whole first BZ as shown in Fig. 2(b). Note that the crossing between the $p_z$ ($\Lambda_6$) band and the upper $d_{xz}/d_{yz}$ ($\Lambda_7$) band is protected by the crystal symmetry. The dashed circle in Fig. 2(b) shows the band inversions and the SOC gap. As shown in Fig. 2(c), the weights of the Fe $d_{xz}/d_{yz}$, Fe $d_{z^2}$ and As $p_z$ orbitals are proportional to the width of the red, green, and blue curves, respectively. The dispersive band along the Γ-Z direction is composed of the As $p_z$ orbital hybridizing with the Fe $d_{z^2}$ orbital. Further parity analysis shows that both the two Fe $d_{xz}/d_{yz}$ bands have even parities at both the Γ and Z points as shown in Fig. 2(c). Note that there is an exchange between the $Z_6^+$ state of the Fe $d_{z^2}$ orbital character and the $Z_8^-$ state of the As $p_z$ orbital character at the Z point as shown in Fig. 2(c). Therefore, the As $p_z$ band has an odd parity at the Z point. When the As $p_z$ band with odd parity crosses the Fe $d_{xz}/d_{yz}$ bands with even parities along the Γ-Z direction, band inversions with opposite parities occur at the Z point, which leads to nontrivial topological properties. For systems with inversion symmetry, the method proposed by Fu and Kane [67], which is based on the parity eigenvalues of occupied bands at eight time-reversal invariant momenta points, can be applied to compute the topological invariants [68]. By defining the occupied eigenspace as spanned by the states below the blue dashed curve in Fig. 2(b), the Fu-Kane parity criterion is well defined for the occupied bands below the Fermi curve in LiFeAs. The $Z_2$ topological invariants are calculated to be (1;000), which indicates that LiFeAs is in a nontrivial topological phase, in consistent with a previous calculation [30].

We continue to compare the DFT+DMFT band structure [Fig. 2(d)] to the DFT

band structure [Fig. 2(c)] of LiFeAs. Whereas electronic correlation strongly renormalizes the DFT bands of Fe 3$d$ electrons by a factor of ~3.5, the DFT+DMFT band structure shows similar band inversions to the DFT band structure and has the same $Z_2$ topological invariants discussed above. Because electronic correlation compresses the bandwidth, the SOC gap between the $p_z$ ($\Lambda_6$) band and the lower $d_{xz}/d_{yz}$ ($\Lambda_6$) band along the $\Gamma$-$Z$ direction is ~24 meV in the DFT band structure, whereas it decreases to ~11 meV in the DFT+DMFT band structure.

To further confirm the nontrivial topological properties of iron-based superconductors, we calculate the surface states on the (001) surface. Based on the DFT band structure of LiFeAs, the calculated surface states shown in Fig. 2(e) have Dirac-cone-type topological surface states at the $\bar{\Gamma}$ point, which is consistent with a previous calculation [30]. The topological surface states include a TI Dirac cone and a Dirac cone above the TI Dirac cone, termed the topological Dirac semimetal (TDS) state. The avoided crossing between the As $p_z$ band and the lower Fe $d_{xz}/d_{yz}$ band leads to the lower TI Dirac cone, whereas the protected crossing between the As $p_z$ band and the upper Fe $d_{xz}/d_{yz}$ band results in the upper TDS Dirac cone [30].

Considering the strong electronic correlation, the DFT+DMFT surface states [Fig. 2(f)] are computed from the DFT+DMFT tight-binding Hamiltonian obtained by fitting to the DFT+DMFT band structure. Comparing to the DFT topological surface states (Dirac cones) in Fig. 2(e), we find that electronic correlation moves the topological surface states, i.e., the lower TI Dirac cone and the upper TDS Dirac cone, much closer to the Fermi level as shown in Fig. 2(f). Note that the TI Dirac cone is crucial for realizing topological superconductivity and Majorana zero modes, hence, it has to be very close to the Fermi level. We note that the DFT+DMFT TI Dirac cone is extremely close to the experimental result [30], in strong contrast to the DFT TI Dirac cone. In addition, the DFT+DMFT TI Dirac cone is well separated from the bulk states as shown in Fig. 2(f), whereas the DFT TI Dirac cone merges into the bulk states shown in Fig. 2(e).

Similar to LiFeAs, the same band inversions between the As $p_z$ band and the Fe

$d_{xz}/d_{yz}$ bands also occur along the Γ-Z direction in both the DFT and DFT+DMFT band structures of NaFeAs (Figs. S1(a) and S1 (b) in the Supplemental Material [69]), BaFe2As2 [Figs. 3(a) and 3(b)], KFe2As2 (Figs. S2(a) and S2(b) in the Supplemental Material [69]), and SrFe2As2 (Figs. S3(a) and S3(b) in the Supplemental Material [69]). The DFT and DFT+DMFT topological surface states on the (001) surface around the $\bar{\Gamma}$ point in NaFeAs, BaFe2As2, KFe2As2, and SrFe2As2 are shown in Figs. S1(c) and S1(d) in the Supplemental Material [69], Figs. 3(c), 3(d), and S2(c), S2(d), S3(c), and S3(d) in the Supplemental Material [69], respectively. It is evident that all the DFT+DMFT topological surface states (TI Dirac cones and TDS Dirac cones) are much closer to the Fermi level than their DFT counterparts due to electronic correlation effects.

## B. Electronic correlation is crucial for determining the position of the As/Se $p_z$ band

The nontrivial band topology discussed in Sec. III A is due to the band inversions between the As $p_z$ band and the Fe $d_{xz}/d_{yz}$ bands along the Γ-Z direction. Since the Fe $d_{xz}/d_{yz}$ bands could not change their positions much because of the Fe 3$d$ valence states in iron-based superconductors, the position of the As/Se $p_z$ band is crucial for the existence of the aforementioned band inversions and the nontrivial band topology. We show that the position of the As/Se $p_z$ band is very sensitive to both the As/Se height to the nearest Fe plane and electronic correlation. Moreover, electronic correlation is also crucial for determining the As/Se height in iron-based superconductors. Therefore, electronic correlation is very important for determining the position of the As/Se $p_z$ band and the band topology of iron-based superconductors.

We illustrate this in FeSe. Keeping the experimental lattice constants, we optimize the internal coordinates of Se ions using both DFT and DFT+DMFT [66,70]. The DFT and DFT+DMFT optimized Se heights from the nearest Fe plane are 1.331 and 1.496 Å, respectively. Comparing to the experimental Se height of 1.476 Å [59], DFT severely underestimates the Se height by 10%, whereas DFT+DMFT reproduces the experimental Se height very well with 1% error. Therefore, it is important to take into account electronic correlation to obtain the equilibrium Se/As position in iron-based superconductors.

To demonstrate the dependence of the Se $p_z$ band position on the Se height from the Fe plane, we further carry out DFT calculations of the band structure of FeSe using the experimental lattice constants with both the DFT-optimized and the experimental Se heights. With the DFT optimized Se height, the position of the Se $p_z$ band along the Γ-Z direction (centered around 1 eV above the Fermi level) is much higher than the Fe 3$d$ bands around the Fermi level, hence, the aforementioned band inversions are absent as shown in Fig. 4(a). However, when experimental Se height is used, the Se $p_z$ band along the Γ-Z direction is significantly pushed down by about 0.7 eV and crosses the three Fe 3$d$ $t_{2g}$ bands as shown in Fig. 4(b). As a result, the aforementioned band inversions occur at the Z point, which leads to a nontrivial band topology. It indicates that the position of the Se $p_z$ band is very sensitive to Se height and can change the band topology from trivial to nontrivial.

We now show how electronic correlation changes the position of the Se $p_z$ band without changing the Se height by comparing the DFT and DFT+DMFT calculated band structures of FeSe with SOC using the same experimental crystal structure including the Se position. The aforementioned band inversions occur along the Γ-Z direction in both the DFT and DFT+DMFT band structures as shown in Figs. 4(c) and 4(d). The dispersive Se $p_z$ band along the Γ-Z direction has a bandwidth of about 0.38 eV in DFT, whereas it is renormalized by electronic correlation to about 0.2 eV in DFT+DMFT. More importantly, electronic correlation moves the bottom of the dispersive Se $p_z$ band at the Z point from ~0.1 eV above the Fermi level in DFT to ~0.04 eV below the Fermi level in DFT+DMFT. Therefore, even with the same Se height, electronic correlation pushes down the Se $p_z$ band along the Γ-Z direction.

In addition, electronic correlation renormalizes the DFT bandwidth of Fe 3$d$ electrons by a factor of approximately 4.1 as shown in Figs. 4(c) and 4(d). As discussed in Sec. III A, the (avoided) crossings of the Se $p_z$ band and the Fe $d_{xz}/d_{yz}$ bands are much closer to the Fermi level in DFT+DMFT than DFT due to electronic correlation effects. As a result, the calculated topological surface states including the TI Dirac cone and TDS Dirac cone shown in Figs. 4(e) and 4(f) are much closer to the Fermi level in

DFT+DMFT [Fig. 4(f)] than DFT [Fig. 4(e)].

### C. Electronic correlation can induce topological phase transitions

Thanks to the strong electronic correlation effects on the position of the As/Se $p_z$ band, we find that some iron-based superconductors including LaFeAsOH$_{0.2}$ and CaFeAsF exhibiting trivial band topology in DFT calculations become topologically nontrivial in DFT+DMFT calculations. In other words, electronic correlation can change the band topology of some iron-based superconductors by pushing down the As/Se $p_z$ band towards the Fermi level and induce a topological phase transition from trivial band topology in DFT calculations to nontrivial band topology in DFT+DMFT calculations.

We demonstrate that electronic correlation induces a topological phase transition in electron doped LaFeAsO. As shown in Figs. S4(a) and S4(b) in the Supplemental Material [69], there are no band inversions along the Γ-$Z$ direction between the As $p_z$ band and the Fe $d_{xz}/d_{yz}$ bands in either the DFT or DFT+DMFT calculated band structures of the parent compound LaFeAsO. Therefore, the parent compound LaFeAsO has a trivial band topology in both the DFT and DFT+DMFT calculations. With increasing electron doping, the As $p_z$ band along the Γ-$Z$ direction is pushed down towards the Fermi level as can be seen by comparing Figs. 5(a) and 5(b) to Figs. S4(a) and S4(b) in the Supplemental Material [69]. At 0.2 electrons doping, there are still no aforementioned band inversions along the Γ-$Z$ direction in the DFT band structure of LaFeAsOH$_{0.2}$ [Fig. 5(a)], and no topological surface states appear on the (001) surface as shown in Fig. 5(c). However, in the DFT+DMFT band structure of LaFeAsOH$_{0.2}$ [Fig. 5(b)], it can be seen that the As $p_z$ band along the Γ-$Z$ direction is further pushed down by electronic correlation and crosses the three Fe 3$d$ $t_{2g}$ bands, which leads to the same band inversions and parity exchange at the $Z$ point as in LiFeAs discussed in Sec. III A. The corresponding topological surface states including the TI Dirac cone and TDS Dirac cone on the (001) surface are shown in Fig. 5(d), confirming LaFeAsOH$_{0.2}$ is in a nontrivial topological phase when electronic correlation is considered. It indicates that electronic correlation induces a topological phase transition in H-doped

LaFeAsO with increasing doping from zero to 0.2 H per Fe.

Similarly, we find that the DFT band structure of CaFeAsF is topologically trivial [Fig. 6(a)] with no topological surface states on the (001) surface [Fig. 6(c)]. In contrast, when electronic correlation is included, the aforementioned band inversions and parity exchange occur in the DFT+DMFT band structure as shown in Fig. 6(b). The TI Dirac cone and TDS Dirac cone can be readily seen in the calculated surface states on the (001) surface [Fig. 6(d)]. It shows that electronic correlation transforms the electronic structure of CaFeAsF from topologically trivial to topologically nontrivial.

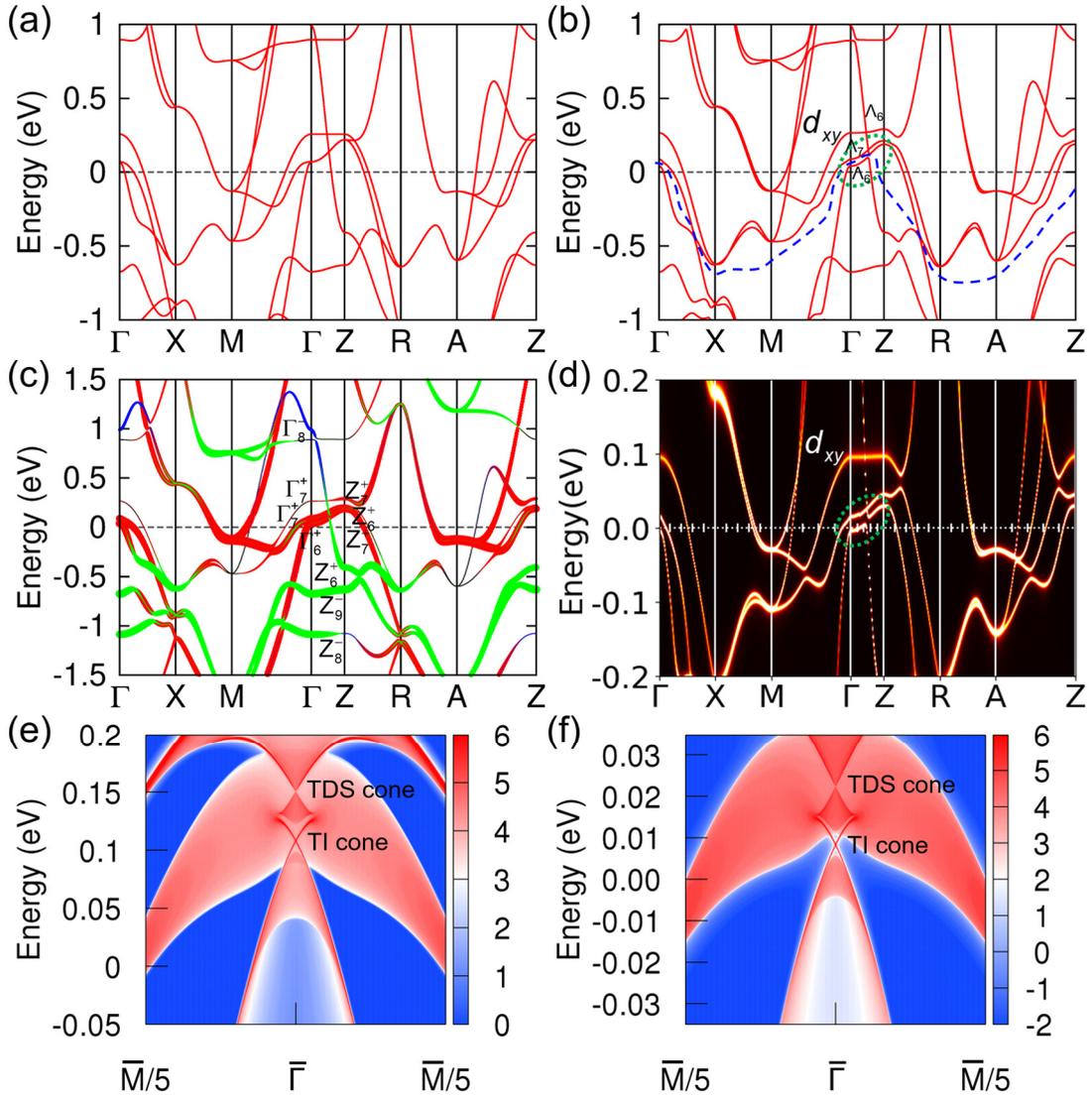

FIG. 2. Band structures and topological surface states of LiFeAs. (a) DFT band

structure without SOC. (b) DFT band structure with SOC. Two $\Lambda_6$ bands hybridize to open a gap along the $\Gamma$-$Z$ line. The dashed circle indicates the band inversions and SOC gap. The blue dashed line indicates the Fermi curve across the SOC gap. (c) The projected band structure with SOC. The weights of the Fe $d_{xz}/d_{yz}$, As $p_z$, and Fe $d_{z^2}$ orbitals are proportional to the width of the red, blue, and green curves, respectively. (d) DFT+DMFT band structure with SOC. The dashed circle marking the same region in (b) shows similar band inversions to those in the DFT band structure. (e) DFT and (f) DFT+DMFT surface states on the (001) surface. The TI Dirac cone and TDS Dirac cone are the topological surface states.

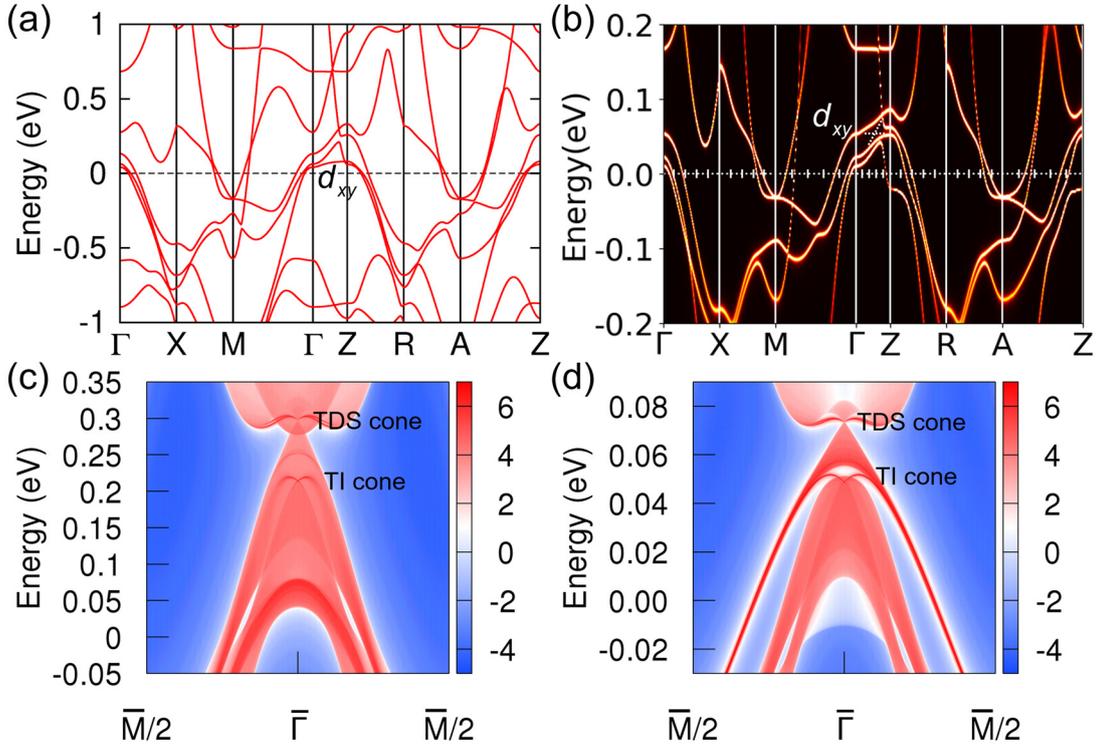

FIG. 3. Band structures and topological surface states of BaFe$_2$As$_2$. (a) DFT band structure with SOC. (b) DFT+DMFT band structure with SOC. The two white dashed lines indicate the Fe $d_{xy}$ band and the upper Fe $d_{xz}/d_{yz}$ band cross and hybridize to open a gap along the $\Gamma$-$Z$ direction. (c) DFT and (d) DFT+DMFT surface states on the (001) surface. The TI Dirac cone and TDS Dirac cone are the topological surface states.

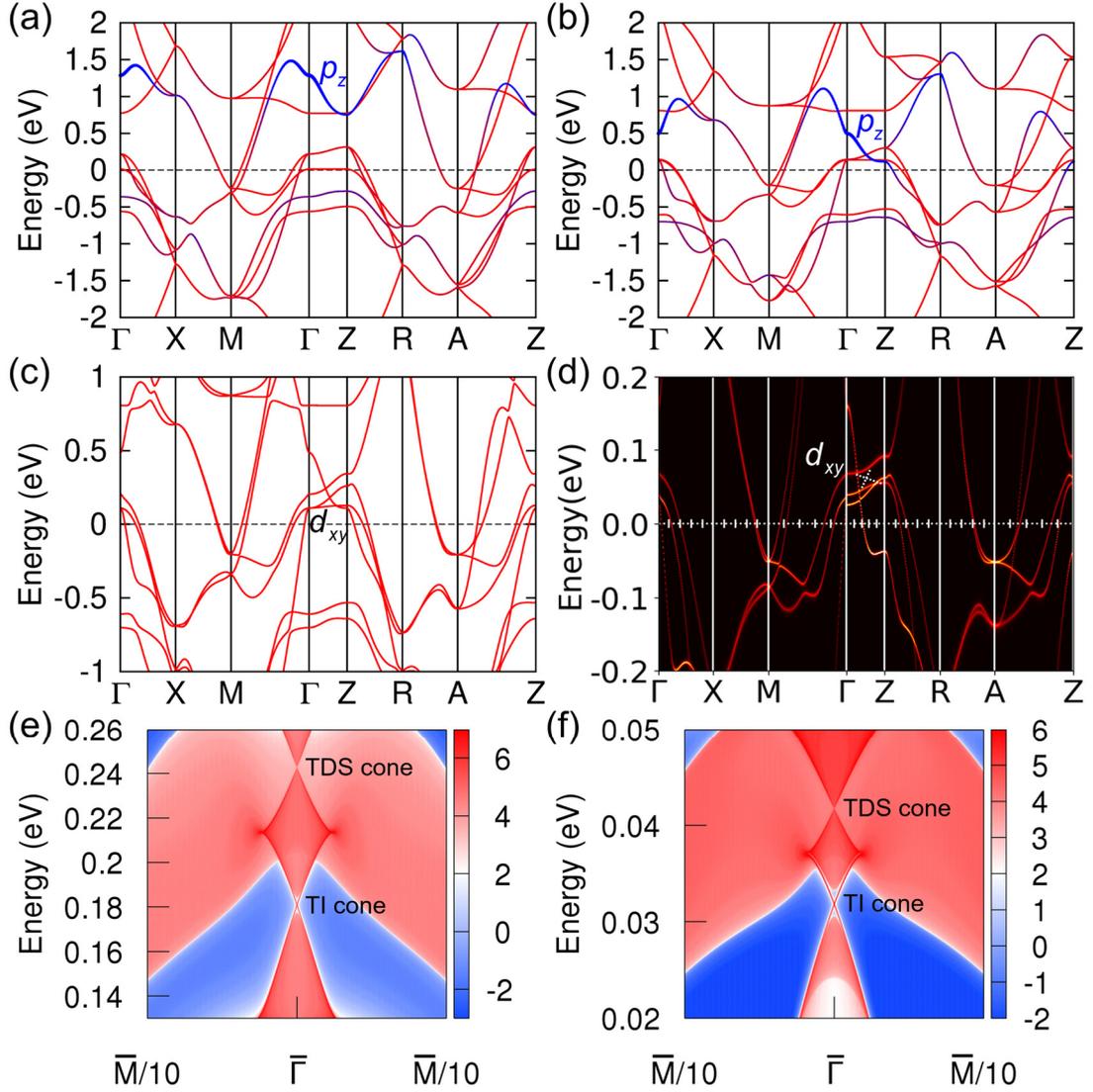

FIG. 4. Band structures and topological surface states of FeSe. (a) and (b) DFT band structures using the DFT optimized Se height (1.331 Å) (a) and the experimental Se height (1.476 Å) (b) without SOC. The weight of the Se $p_z$ orbital is proportional to the width of the blue curves in (a) and (b). (c) DFT and (d) DFT+DMFT band structures using the experimental Se height (1.476 Å) with SOC. The two white dashed lines in (d) indicate the Fe $d_{xy}$ band and the upper Fe $d_{xz}/d_{yz}$ band cross and hybridize to open a gap along the Γ-Z direction. (e) DFT and (f) DFT+DMFT surface states on the (001) surface based on the band structures in (c) and (d), respectively. The TI Dirac cone and TDS Dirac cone are the topological surface states.

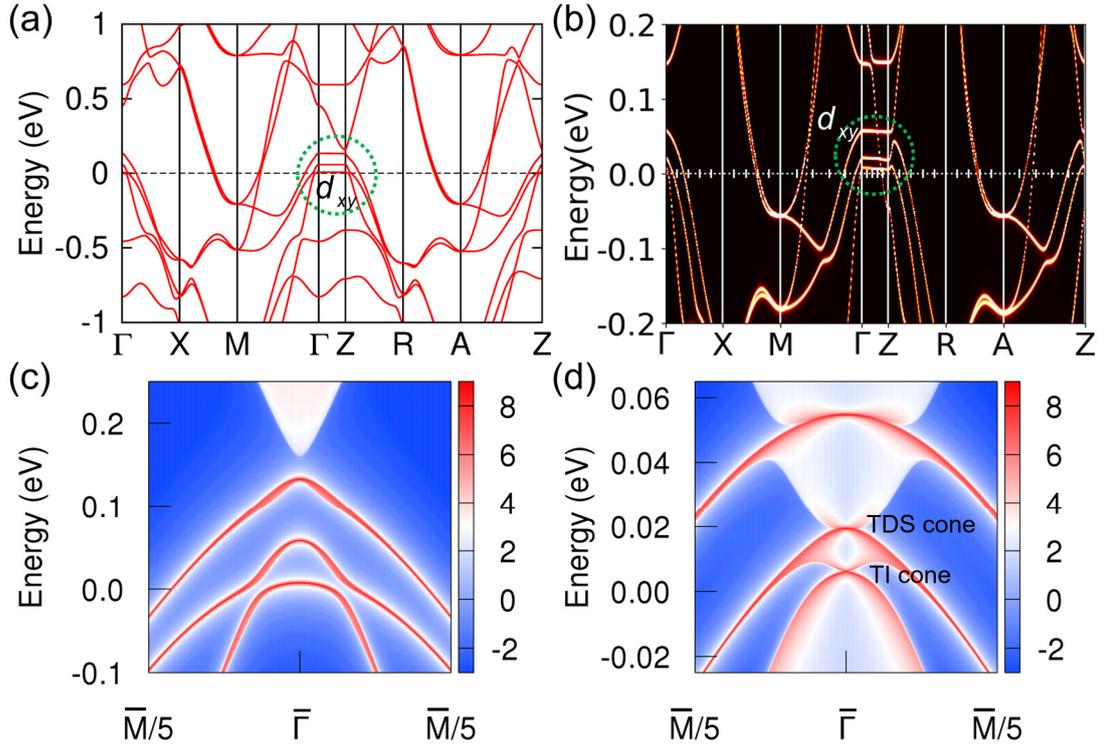

FIG. 5. Band structures and topological surface states of LaFeAsOH$_{0.2}$. (a) DFT band structure with SOC. The dashed circle shows that the As $p_z$ band does not cross the Fe $d_{xz}/d_{yz}$ bands along the Γ-Z direction. No band inversion occurs along the Γ-Z direction. (b) DFT+DMFT band structure with SOC. The dashed circle marks the same region in (a), but there are band inversions along the Γ-Z direction. (c) DFT and (d) DFT+DMFT surface states on the (001) surface. No topological surface state exists in (c) in the DFT calculation. The TI Dirac cone and TDS Dirac cone in (d) are the topological surface states in the DFT+DMFT calculation.

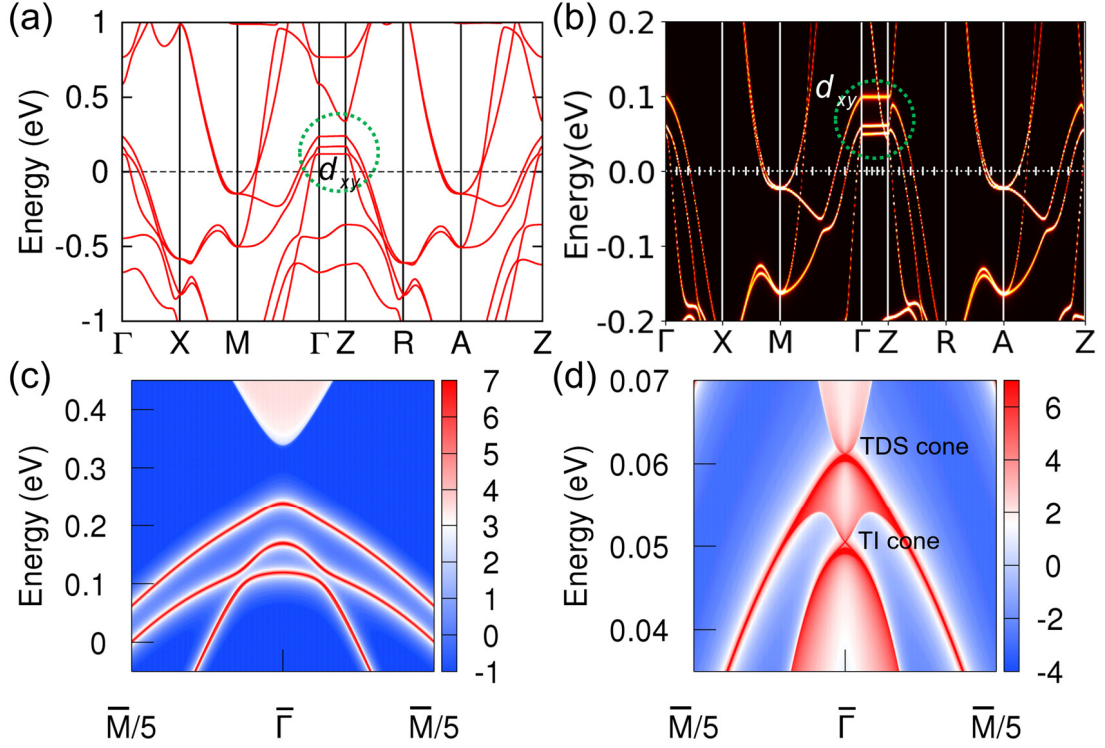

FIG. 6. Band structures and topological surface states of CaFeAsF. (a) DFT band structure with SOC. The dashed circle shows that the As $p_z$ band does not cross the Fe $d_{xz}/d_{yz}$ bands along the Γ-Z direction. No band inversion occurs along the Γ-Z direction. (b) DFT+DMFT band structure with SOC. The dashed circle marks the same region in (a), but there are band inversions along the Γ-Z direction. (c) DFT and (d) DFT+DMFT surface states on the (001) surface. No topological surface state exists in (c) in the DFT calculation. The TI Dirac cone and TDS Dirac cone in (d) are the topological surface states in the DFT+DMFT calculation.

## IV. CONCLUSIONS

In conclusion, we use DFT+DMFT to systematically study the band topology and topological surface states of iron-based superconductors. We find that the strong electronic correlation of Fe $3d$ electrons plays an important role in determining the band topology and topological surface states of many iron-based superconductors. Electronic correlation not only renormalizes the Fe $3d$ bands, but also renormalizes and pushes down the As/Se $p_z$ band along the Γ-Z direction. As a result, electronic correlation moves the topological surface states including the TI Dirac cone and the TDS Dirac cone much closer to the Fermi level than those obtained in DFT calculations in all the iron-based superconductors we study. The closeness of the topological surface states to the Fermi level is important for realizing topological superconducting surface states and Majorana zero modes. More interestingly, electronic correlation plays a crucial role in determining the band topology of iron-based superconductors: it induces a topological phase transition in H-doped LaFeAsO and transforms the topologically trivial DFT band structure of CaFeAsF into a topologically nontrivial band structure. As a result, it creates TI-Dirac-cone and TDS-Dirac-cone topological surface states on the (001) surface that are nonexisting in DFT calculations. Our paper demonstrates that properly treating the strong electronic correlation is important for accurately computing the band topology and topological surface states of iron-based superconductors and other strongly correlated materials.

## ACKNOWLEDGMENTS

This work was supported by the National Natural Science Foundation of China (Grants No. 12074041 and 11674030), the Foundation of the National Key Laboratory of Shock Wave and Detonation Physics (Grant No. 6142A03191005), the National Key Research and Development Program of China through Contract No. 2016YFA0302300, and the start-up funding of Beijing Normal University. The calculations were carried out with high performance computing cluster of Beijing Normal University in Zhuhai.

Supplemental Material for "Correlation-corrected band topology and topological surface states in iron-based superconductors"

Xiaobo Ma,[#] Guangwei Wang,[#] Rui Liu, Tianye Yu, Yiran Peng, Pengyu Zheng and Zhiping Yin*

*Department of Physics and Center for Advanced Quantum Studies, Beijing Normal University, Beijing 100875, China*

In the Supplemental Material, we present the band structures and surface states for NaFeAs, $KFe_2As_2$, and $SrFe_2As_2$ based on DFT and DFT+DMFT calculations, as shown in Figs. S1-S3. We also provide the DFT and DFT+DMFT band structures of the parent compound LaFeAsO, as shown in Figs. S4(a) and S4(b).

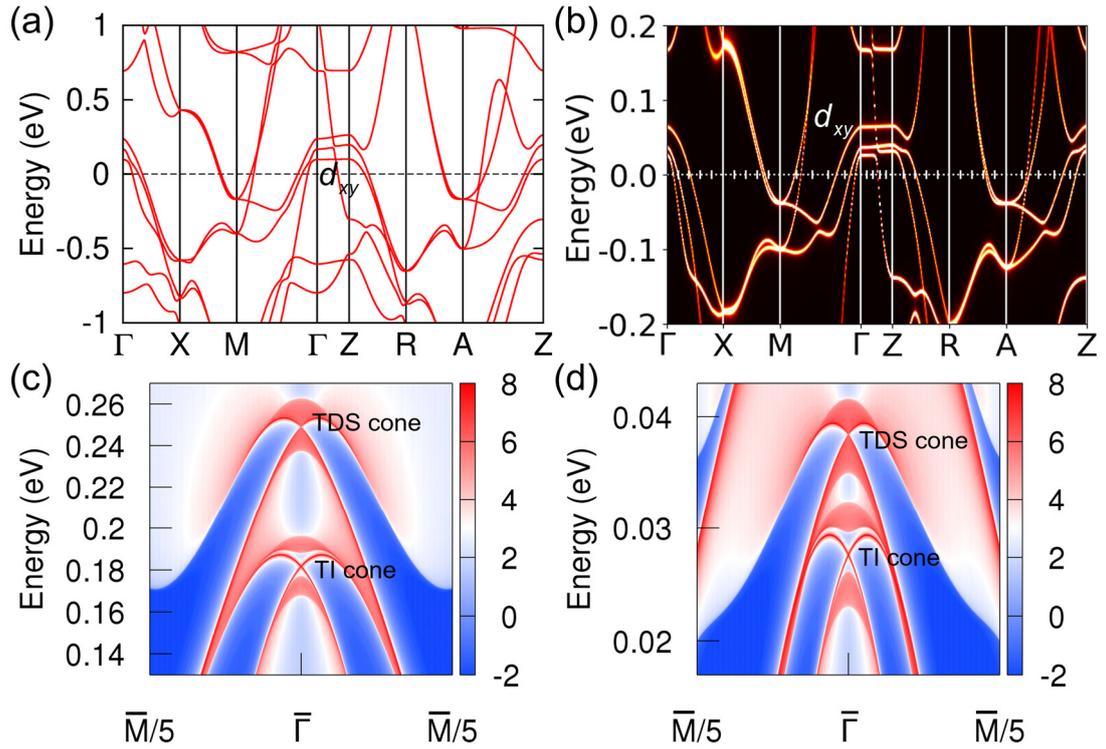

FIG. S1. Band structures and topological surface states of NaFeAs. (a) DFT band structure with SOC. (b) DFT+DMFT band structure with SOC. (c) DFT and (d) DFT+DMFT surface states on the (001) surface. The TI Dirac cone and TDS Dirac cone are the topological surface states.

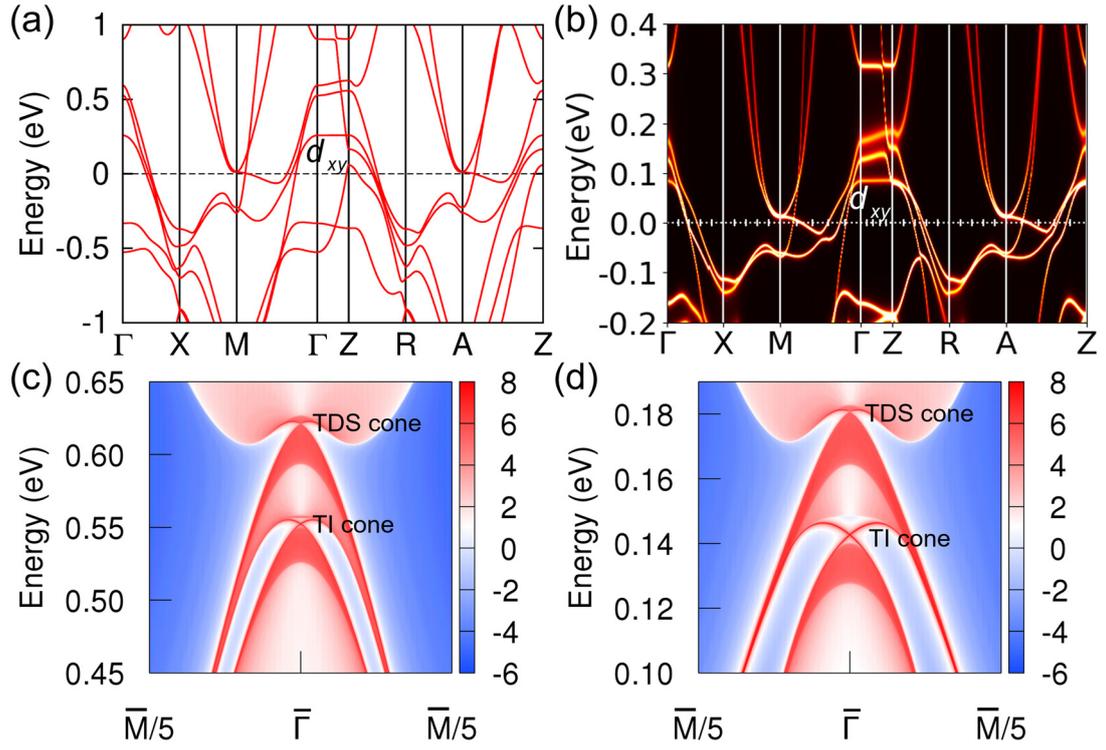

FIG. S2. Band structures and topological surface states of $KFe_2As_2$. (a) DFT band structure with SOC. (b) DFT+DMFT band structure with SOC. (c) DFT and (d) DFT+DMFT surface states on the (001) surface. The TI Dirac cone and TDS Dirac cone are the topological surface states.

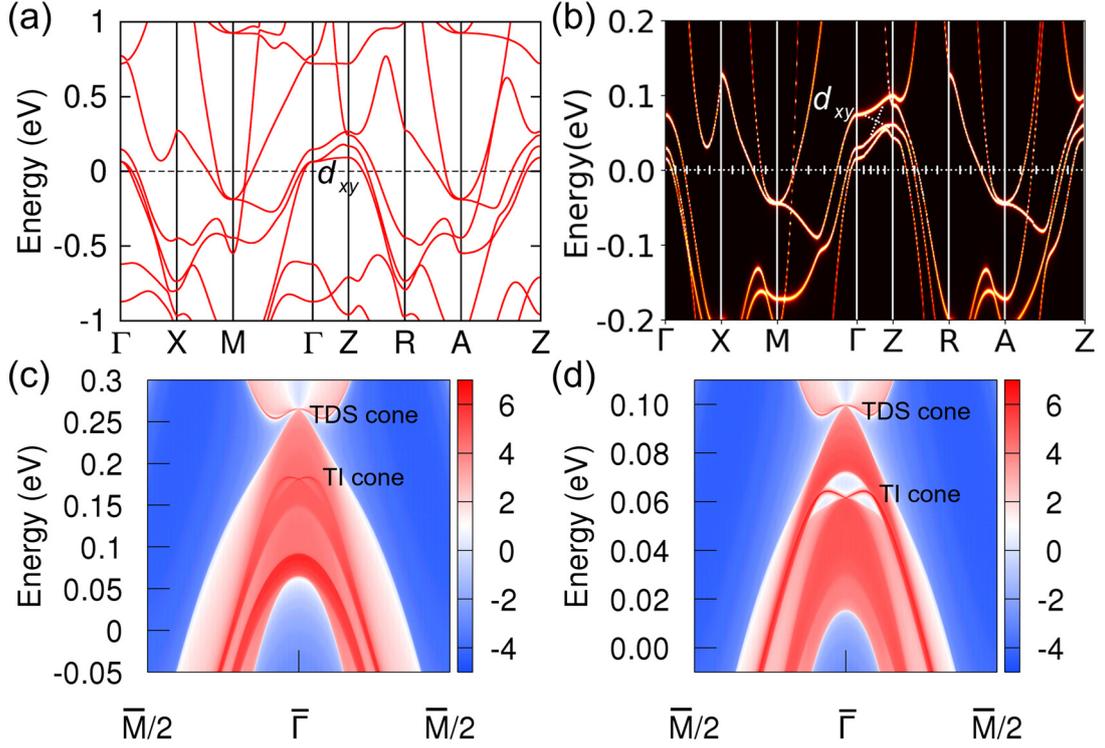

FIG. S3. Band structures and topological surface states of SrFe$_2$As$_2$. (a) DFT band structure with SOC. (b) DFT+DMFT band structure with SOC. The two white dashed lines indicate the Fe $d_{xy}$ band and the upper Fe $d_{xz}/d_{yz}$ band cross and hybridize to open a gap along the Γ-Z direction. (c) DFT and (d) DFT+DMFT surface states on the (001) surface. The TI Dirac cone and TDS Dirac cone are the topological surface states.

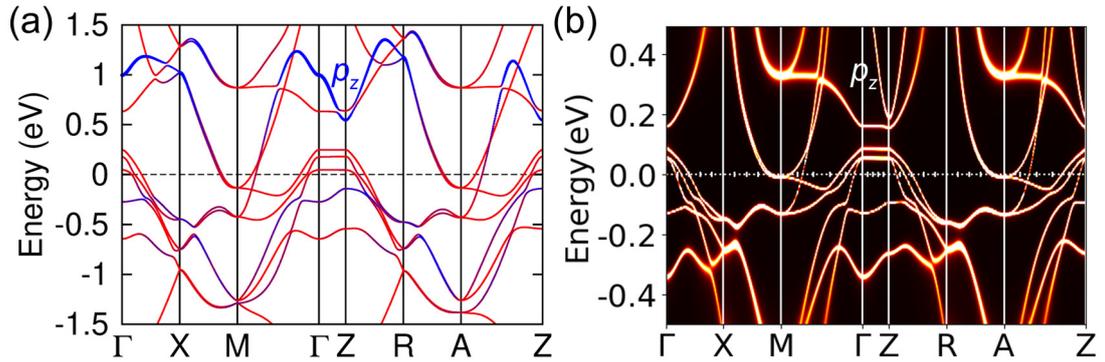

FIG. S4. Band structures of LaFeAsO. (a) DFT band structure with SOC. (b) DFT+DMFT band structure with SOC. The As $p_z$ band does not cross the Fe $d_{xz}/d_{yz}$ bands and no band inversion occurs along the Γ-Z direction in both (a) and (b). The weight of the As $p_z$ orbital is proportional to the width of the blue curves in (a).